\documentclass[twocolumn,showpacs,aps,prl]{revtex4-1}
\usepackage{amsmath,amssymb,graphicx,hyperref}

\newcommand{\beq}{\begin{eqnarray}}
\newcommand{\eeq}{\end{eqnarray}}
\newcommand{\beqq}{\begin{eqnarray*}}
\newcommand{\eeqq}{\end{eqnarray*}}

\usepackage{graphicx}
\usepackage{dcolumn}
\usepackage{bm}
\usepackage{color}
\usepackage{amsmath,amssymb,amsthm}
\usepackage{hyperref}

\begin{document}

\title{Self-Learning Monte Carlo Method in Fermion Systems}

\author{Junwei Liu$^{1\dagger}$, Huitao Shen$^{1\dagger}$, Yang Qi$^{1}$, Zi Yang Meng$^{2}$ and Liang Fu$^{1}$}
\email{liangfu@mit.edu\\$^{\dagger}$The first two authors contributed equally to this work.}
\affiliation{$^1$Department of physics, Massachusetts Institute of Technology, Cambridge, MA 02139, USA}
\affiliation{$^2$Institute of Physics, Chinese Academy of Sciences, Beijing 100190, China}

\date{\today}

\begin{abstract}
We develop the self-learning Monte Carlo (SLMC) method, a general-purpose numerical method recently introduced to simulate many-body systems, for studying interacting fermion systems. Our method uses a highly-efficient update algorithm, which we design and dub ``cumulative update'', to generate new candidate configurations in the Markov chain based on a self-learned bosonic effective model. From general analysis and numerical study of the double exchange model as an example, we find the SLMC with cumulative update drastically reduces the computational cost of the simulation, while remaining statistically exact. Remarkably, its computational complexity is  far less than the conventional algorithm with local updates.
\end{abstract}


\maketitle

Monte Carlo (MC) method is an unbiased numerical tool that obtains statistically exact results by sampling configurations according to a probability distribution \cite{Metropolis1953,HASTINGS1970,Binder1995,NewmanBarkema1999,Gubernatis2003,LandauBinder2005,Sandvik2010}. Configurations may be generated sequentially through the reversible Markov process obeying the detailed balance principle (DBP). In order for a MC simulation to be efficient, the process of generating a new configuration from the current one should be fast, and consecutive configurations should be uncorrelated. The performance of conventional MC method is severely impeded when either of the two conditions fails.

Recently, we introduced a new method dubbed ``self-learning Monte Carlo'' (SLMC), to speed up configuration updates in MC simulations \cite{liu2016self}. SLMC consists of two stages: learning and simulating. First, we perform trial simulations with the conventional local update method to generate a large set of configurations along with their weights, which serve as the training data. By fitting the configuration distribution, we learn an effective Hamiltonian $H_{\rm eff}$ for the system, which can be simulated faster than the original Hamiltonian $H$. Next, in performing the actual simulation,  we use $H_{\rm eff}$ to propose smart moves in configuration space. The acceptance of proposed moves is set properly to satisfy the detailed balance condition of the original Hamiltonian, ensuring the simulation is statistically exact.

SLMC is a general-purpose method rooted in the philosophy ``first learn, then earn''. In our previous work \cite{liu2016self}, SLMC is implemented for classical statistical models near second-order phase transitions. In such systems, conventional MC simulation suffers from critical slowing down because successive configurations are highly correlated \cite{SwendsenWang1987,Wolff1989}. We showed that SLMC method can significantly reduce the autocorrelation time. As an example, SLMC simulation on a generalized Ising model is found to be 10-20 times faster.

In this work, we develop a generic SLMC method for simulating interacting fermions as well as mixed Bose-Fermi systems, and demonstrate its enormous power. By theoretical analysis and numerical simulation, we show that SLMC method {\it generally} reduces the complexity of simulating fermion systems, thus achieving a tremendous speedup that {\it grows} with system size. The central component of our method is a highly efficient ``cumulative update'' algorithm, which we introduce for updating field configurations to which fermions couple. 

Below we first present SLMC method with cumulative update in full generality and theoretically analyze its complexity, i.e., determine the scaling of the computational cost with system size. Next, we apply our method to the double exchange model, which describes itinerant electrons interacting with localized spins. Our method achieves a speedup of at least $O(L^{d})$, where $L$ is the system size and $d$ is the spatial dimension. For the double exchange model on $8\times 8 \times 8$ cubic lattice, SLMC simulation is numerically shown to be over $10^3$ times faster than conventional method.

In quantum MC simulations of interacting fermion systems, we may employ the Hubbard-Stratonovich transformation \cite{stratonovich1957,hubbard1959} to write the partition function in the form of fermions coupled to fluctuating fields \cite{blankenbecler1981,hirsch1985,hirsch1986,white1989}
\begin{eqnarray}
  \label{eq:Z}
  Z=\sum_{\phi(\tau)}\text{det}\left[
  \mathbf{I}+ \prod_{\tau} e^{-\Delta \tau H_f[\phi(\tau)]}\right] \equiv \sum_{\phi(\tau)} W[\phi(\tau)],
\end{eqnarray}
where $\phi$ is a space- and time-dependent field, $H_f[\phi]$ is the single-particle Hamiltonian of fermions moving in the background of the field $\phi$, $\Delta \tau= \beta/N$ is the duration of time slice in Trotter decomposition \cite{trotter1959,suzuki1976} and $\tau=n\Delta \tau$ with $n=1, ..., N$. In the MC simulation, we sample configurations of $\phi$ with weights $W[\phi]$, which is determined by integrating out fermion degrees of freedom. Partition functions of the form Eq.\eqref{eq:Z} also appear in systems of fermions interacting with dynamical boson fields such as spins or phonons \cite{Erez2012,Lederer2015,li2015a,Schattner2016prl,Schattner2016prx,xu2016arXiv}. 

The partition function Eq.\eqref{eq:Z} can also describe systems of fermions coupled to classical spins or other classical degrees of freedom. A well-known example is the double exchange model \cite{zener1951,anderson1955,de1960} describing itinerant electrons coupled to a lattice of localized spins, which are represented by classical vectors of unit length.
In such cases, the field $\phi$ in Eq.\eqref{eq:Z} is space-dependent but {\it time-independent}.  The partition function Eq.\eqref{eq:Z} then simplifies to
\beq
Z =\sum_{\phi}\text{det}\left[
  \mathbf{I}+  e^{-\beta H_f[\phi]}\right] \equiv \sum_{\phi} W[\phi]. \label{Z'}
\eeq

Computing the weight $W$ in both Eq.\eqref{eq:Z} and Eq.\eqref{Z'} involves calculating the fermion determinant.  
This task is very time-consuming and its computational cost grows polynomially with system size, which is a major bottleneck of MC simulations in fermion systems.


%

We leave the SLMC treatment of time-dependent fields to a forthcoming work\cite{xu2016}, and from now on, focus on the models of fermions coupled to classical fields, where $\phi$ is time-independent. In this case, the weight $W$ in Eq.\eqref{Z'} can be calculated as $\prod_{n} (1+e^{-\beta E_n(\phi)})$, where $\{E_n(\phi)\}$ is the single-particle energy spectrum obtained by the exact diagonalization (ED) of the fermion Hamiltonian $H_f[\phi]$ for a given $\phi$ configuration. Performing this ED has the complexity of $O(L^{3d})$. (Algorithms that compute $W[\phi]$ approximately can be faster \cite{motome1999,alonso2001,alvarez2005,Kumar2006,Jeroen2010,mukherjee2015}.) In the conventional MC simulation using the Metropolis algorithm, the ED needs to be performed every time $\phi$ is updated on a single site. Therefore, the computational cost for each full sweep of the lattice in the simulation grows as $O(L^{4d})$. Assuming the configurations become uncorrelated after $\tau_0$ iterations of such sweeps, the total cost of generating two successive, statistically independent configurations is $O(\tau_0 L^{4d} )$.

In contrast, in SLMC simulation, we first generate a set of configurations using the local update, and fit their weights using an effective model $H_{\rm eff}[\phi]$ for the field $\phi$ such that
\begin{eqnarray}
W[\phi] \simeq e^{-\beta H_{\rm eff}[\phi]}.
\end{eqnarray}
Typically we take $H_{\rm eff}[\phi]$ to have the form of a power series of $\phi$, and determine the coefficients from multi-linear fitting.
$H_{\rm eff}[\phi]$ is to be viewed as an approximation to the exact Hamiltonian for the $\phi$-field after integrating out the fermions $H[\phi] \equiv -\frac{1}{\beta}\ln W[\phi]$, because by construction the Boltzmann distribution of $H_{\rm eff}[\phi]$ approximately reproduces the desired distribution $W[\phi]$ for {\it dominant} $\phi$-field configurations.
Importantly, for a given $\phi$, evaluating $H_{\rm eff}[\phi]$ whose explicit expression has been learned from the fitting is much faster than numerically computing $W[\phi]$ exactly.

Next, in performing the actual simulation, we design a  ``cumulative update'' algorithm to update the $\phi$-field configurations efficiently by the guidance of the effective model $H_{\rm eff}[\phi]$. Starting from the last configuration reached in the Markov chain of $ H[\phi] $, denoted by $\phi_A$, we propose a global move by performing a sequence of local updates as one would do in simulating the effective model $H_{\rm eff}[\phi]$: each local move attempts to change the value of $\phi$ on a random site, and its acceptance probability is determined by the detailed balance condition of $H_{\rm eff}[\phi]$. A sequence of such local updates is performed to generate a new field configuration $\phi_B$ that is sufficiently uncorrelated with $\phi_A$. Then we propose $\phi_B$ as the candidate configuration for the next state in the Markov chain of $ H[\phi] $. As we will describe below, the probability of accepting this move $\phi_A \rightarrow \phi_B$ should be designed properly to ensure that SLMC method is statistically exact.

The advantage of the cumulative update algorithm comes from replacing many iterations of ``expensive'' local updates according to the exact weight $W[\phi]$, which is necessary to obtain two statistically independent configurations in the conventional MC simulation, with cumulative local updates according to the approximate weight $e^{- \beta H_{\rm eff}[\phi]}$, which is numerically much faster.

We now derive the desired probability of accepting the proposed move $\phi_A \rightarrow \phi_B$ in a single step of the cumulative update. As in the general Metropolis-Hastings algorithms \cite{HASTINGS1970}, our update scheme consists of  two stages: first, a candidate configuration $\phi_B$ is generated from $\phi_A$ through a sequence of local updates; second, the update $\phi_A\rightarrow \phi_B$ is accepted with a probability $p(A\rightarrow B)$. The Markov-chain transition probability $P(A\rightarrow B)$ is the product of the probability of generating the particular configuration $\phi_B$ among all the possibilities in the first stage, denoted by $\mathcal S(A\rightarrow B)$, and the probability of accepting $\phi_B$, $p(A\rightarrow B)$. The detailed balance condition for $P(A\rightarrow B)$ requires
\begin{equation}
\frac{P(A\to B)}{P(B\to A)}=\frac{\mathcal{S}(A\to B)}{\mathcal{S}(B\to A)}\frac{p(A\to B)}{p(B\to A)}=\frac{W(B)}{W(A)}, \label{dbp}
\end{equation}
where $ W(A) \equiv W[\phi_A]$ is the exact weight of the configuration $\phi_A$.  Moreover, by construction,
the ratio ${\mathcal{S}(A\to B)}/{\mathcal{S}(B\to A)}$ is set by the detailed balance condition of the effective model
\begin{widetext}
\begin{equation}
\frac{\mathcal{S}(A\to B)}{\mathcal{S}(B\to A)}=\prod_{i=0}^{l_c-1} \frac{\tilde{P}(C_i\to C_{i+1})}{\tilde{P}(C_{i+1}\to C_i)}
=\prod_{i=0}^{l_c-1} e^{-\beta (\tilde{E}_{i+1}-\tilde{E}_{i})}=e^{-\beta(\tilde{E}_B-\tilde{E}_A)}, \label{a}
\end{equation}
\end{widetext}
where $ \phi_{C_0} \equiv \phi_A $ and $ \phi_{C_n} \equiv \phi_B $. $ \tilde{P}( C_i \to C_{i+1}) $ is the transition probability for $i$-th local update of the effective Hamiltonian $H_{\rm eff}[\phi]$, and $\tilde{E}_i \equiv H_{\rm eff}(\phi_{C_i})$ is the energy of the configuration $\phi_{C_i}$ in the effective model. Combining Eqs.\eqref{dbp} and \eqref{a}, we find the desired probability of accepting the candidate configuration found through cumulative update,
\begin{equation}
p(A \to B)=\min\{1,e^{-\beta(E_B-\tilde{E}_B)-(E_A-\tilde{E}_A)}\}, \label{p}
\end{equation}
where $E$ is the energy of the exact Hamiltonian $H[\phi]$. In the ideal case when $ H_{\rm eff}[\phi]= H[\phi]$, $p=1$.

The computational cost of one complete cumulative update step contains two parts: (1) the cost of local updates based on $H_{\rm eff}$, which generally has the complexity $O(l_c)$, where $l_c$ denotes the number of local updates; (2) the cost of ED to compute the acceptance probability in Eq.\eqref{p}, which has the complexity $O(L^{3d})$. Hence, the total cost is $O(l_c)+O(L^{3d})$. In order to obtain two uncorrelated configurations in a single cumulative update step, we take $l_c$ to be of the order of $\tau_0 L^d$, with $\tau_0$ to be the autocorrelation time (a unit $ \tau_0 $ corresponds to a single full sweep of $ L^d $ sites). For systems away from critical points, $\tau_0$ is taken as a constant, and close to critical points $\tau_0$ scales as $L^z$ with $z$ empirically found to be around $2$ \cite{SwendsenWang1987,Wolff1989}.
Thus, the cost of the ED operation almost always dominates in the cumulative update process.
Assuming that the effective model is accurate enough such that the proposed global move $\phi_A\rightarrow \phi_B$ is accepted with a probability of the order of $100\%$, a cumulative update generates a statistically independent configuration with a computational cost of a {\it single} ED step, i.e. $O(L^{3d})$. Compared to the complexity of $O(\tau_0 L^{4d})$ for the conventional local update, this gives a speedup of $O(\tau_0 L^{d}) = O(L^{d+\eta})$ where $\eta=0$ for non-critical systems and $\eta = z$ for critical systems.

We notice that there are faster methods that evaluate the weight in Eq.\eqref{Z'} {\it approximately} without performing an ED \cite{motome1999,alonso2001,alvarez2005,Kumar2006,Jeroen2010,mukherjee2015}. These methods are not exclusive from the SLMC methods, and by incorporating these approximations, we can further reduce the complexity of SLMC.


In the rest of this work, we numerically demonstrate the power of SLMC by simulating the double exchange model \cite{zener1951,anderson1955,de1960}:
\begin{equation}
\hat{H} = -t\sum_{\langle ij \rangle, \alpha} (\hat{c}_{i\alpha}^\dag \hat{c}_{j\alpha}+\mathrm{h.c.})-\frac{J}{2}\sum_{i, \alpha, \beta}\vec{S}_i\cdot  \hat{c}_{i\alpha}^\dag
\vec{\sigma}_{\alpha\beta}\hat{c}_{i\beta},
\label{h}
\end{equation}
where $\langle ij\rangle $ is the summation over the nearest-neighbor pairs of sites. $ \hat{c}_{i\sigma} $ is the fermion annihilation operator of spin $ \sigma $ at site $ i $, and $\vec{\sigma}$ are the Pauli matrices for the fermion spin operator. $\vec{S}_i$ is a classical vector of unit length on site $i$, representing the localized spin that couples to the fermion on the same site. Despite the absence of direct interactions in this model between localized spins on different sites, the presence of itinerant fermions effectively generates a RKKY-type \cite{ruderman1954,kasuya1956,yosida1957} interaction $g_{ij} \vec{S}_i \cdot \vec{S}_j$, where $g_{ij}$ decays with the distance between site $i$ and $j$.

Below we simulate the double exchange model on the three-dimensional (3D) cubic lattice. Previous numerics have shown that for $J/t=16$ and at filling $\nu=1/4$, this model exhibits a phase transition from paramagnet to ferromagnet at a critical temperature $T_c/t\sim 0.12$ \cite{alvarez2005}. We focus on simulations around $T_c$, where the speedup of SLMC is expected to be maximum.

\begin{figure}[tbp]
\includegraphics[width=\columnwidth]{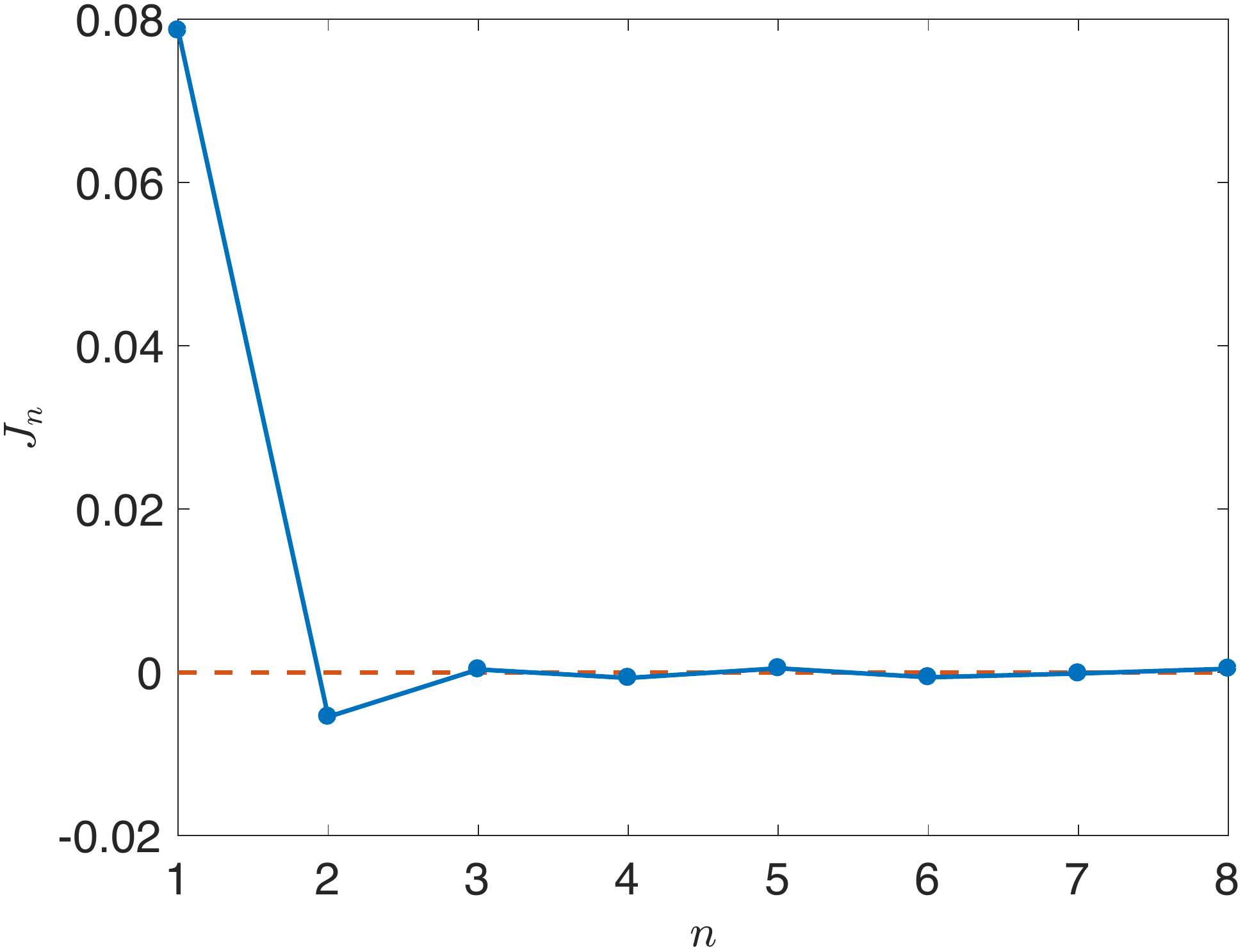}
\caption{(color online) The trained parameters $ J_n $ for the effective model in Eq.\eqref{heff} for $L=4$. The error bars are smaller than the symbol of the data points.
}
\label{fig1}
\end{figure}

Following the general procedure of the SLMC, we first perform trial simulations to train an effective model that includes {\it all} two-body interactions between two localized spins, preserving translational and spin-rotational symmetries
\beq
\label{heff}
H_{\rm eff}=E_0-J_1\sum_{\langle ij\rangle_1} \vec{S}_i\cdot \vec{S}_j-J_2\sum_{\langle ij\rangle_2} \vec{S}_i\cdot \vec{S}_j-\cdots,
\eeq
where $\langle ij\rangle_n $ is the summation over $n$-th nearest-neighbor pairs of localized spins. Denoting $ C_n^\alpha=\sum_{\langle ij\rangle_n}\vec{S}_i\cdot\vec{S}_j$, we train $ E_0 $ and $ J_n $ through a multi-linear regression with $ E^\alpha=E_0+\sum_{n}J_n C_n^\alpha$. To speed up and improve the fitting, we have exploited the reinforcement learning strategy here, i.e., we use the SLMC method on the trained model to generate more uncorrelated spin configurations efficiently, and then further optimize parameters in the effective model using these new training data. This reinforced learning process can be iterated a few times until the desired parameters converge.

\begin{figure}[tbp]
\includegraphics[width=\columnwidth]{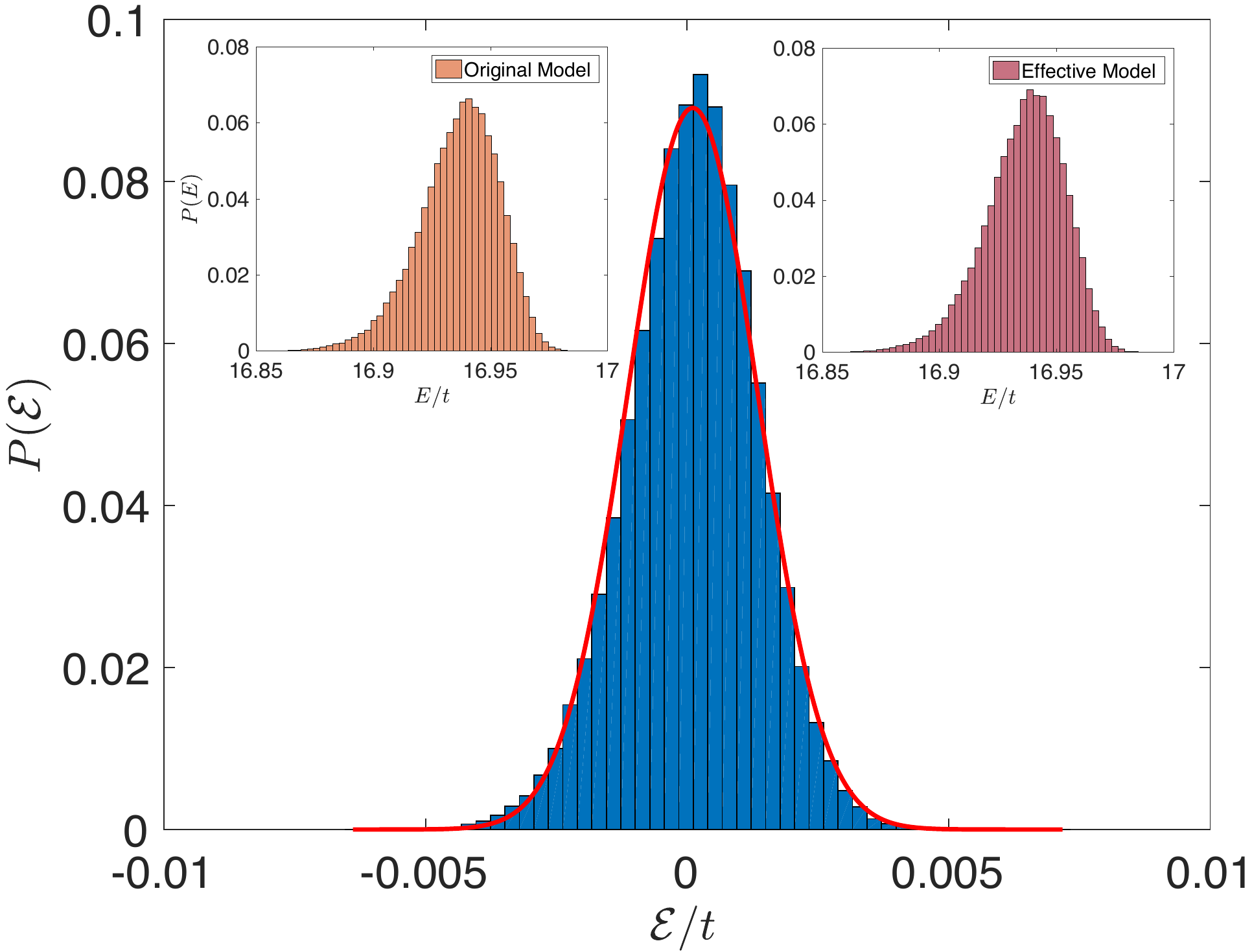}
\caption{(color online) For $ 10^5 $ independent spin configurations on the Markov chain of $ H $, the blue histogram is the distribution of the energy difference $ {\cal E}(S) $ of these configurations between the original model $ H(S) $ and the effective model $ H_{\rm eff}(S) $. The red curve is the fitted Gaussian distribution with $ \sigma = 0.012 $. The upper-left and upper-right insets are distributions of $ E(S) $ and $ E_{\rm eff}(S) $ respectively.
}
\label{fig2}
\end{figure}

The fitting result for $ L=4 $ is shown in the Fig.\ref{fig1}. We see a clear RKKY profile of the spin-spin interaction mediated by itinerant fermions: the coupling strength $ J_n $ decays and oscillates with the distance between two localized spins. To demonstrate how good the fitting is, we sample $ 10^5 $ independent configurations on the Markov chain according to the Boltzmann distribution of the original model and of the effective model. As shown in the insets of Fig.\ref{fig2}, these two Boltzmann distributions look nearly identical.

To quantify the accuracy of the effective model Eq.\eqref{heff}, we measure the differences in the energies of $H_{\rm eff}$ (denoted by $\tilde{E}$) and of $H$ (denoted by $E$) for sample configurations (denoted by $S$) of localized spins drawn from the Boltzmann distribution $e^{-\beta H}$, known as residuals in statistical analysis:
$
{\cal E}_S \equiv  \tilde{E}_S - E_S.
$
As shown in Fig.\ref{fig2}, the distribution of $\cal E$ is Gaussian, with a  peak centered at $ {\cal E} = 0 $. The narrow width of the peak demonstrates the high accuracy of our effective model. In the ideal case where $H_{\rm eff}=H$, the distribution of $\cal E$ becomes a delta function.

We further use the statistical measure $R^2$, also called the coefficient of determination, as a figure of merit (``score'') characterizing how well the effective model replicates the Boltzmann distribution of the original model.
This score is defined through the sum of squares of residuals:
\begin{equation}
R^2 = 1- \frac{\sum_S {\cal E}^2_S }{\sum_S (E_S - \langle E \rangle)^2},
\end{equation}
and ranges from $0$ to $100\%$. A perfect effective model $H_{\rm eff}=H$ has $R^2=100\%$. The scores of our effective model for the double exchange Hamiltonian with $ L=4,6,8 $ are $R^2=99.5\%, 99.8\%, 99.9\%$ respectively.

\begin{figure}[tbp]
\includegraphics[width=\columnwidth]{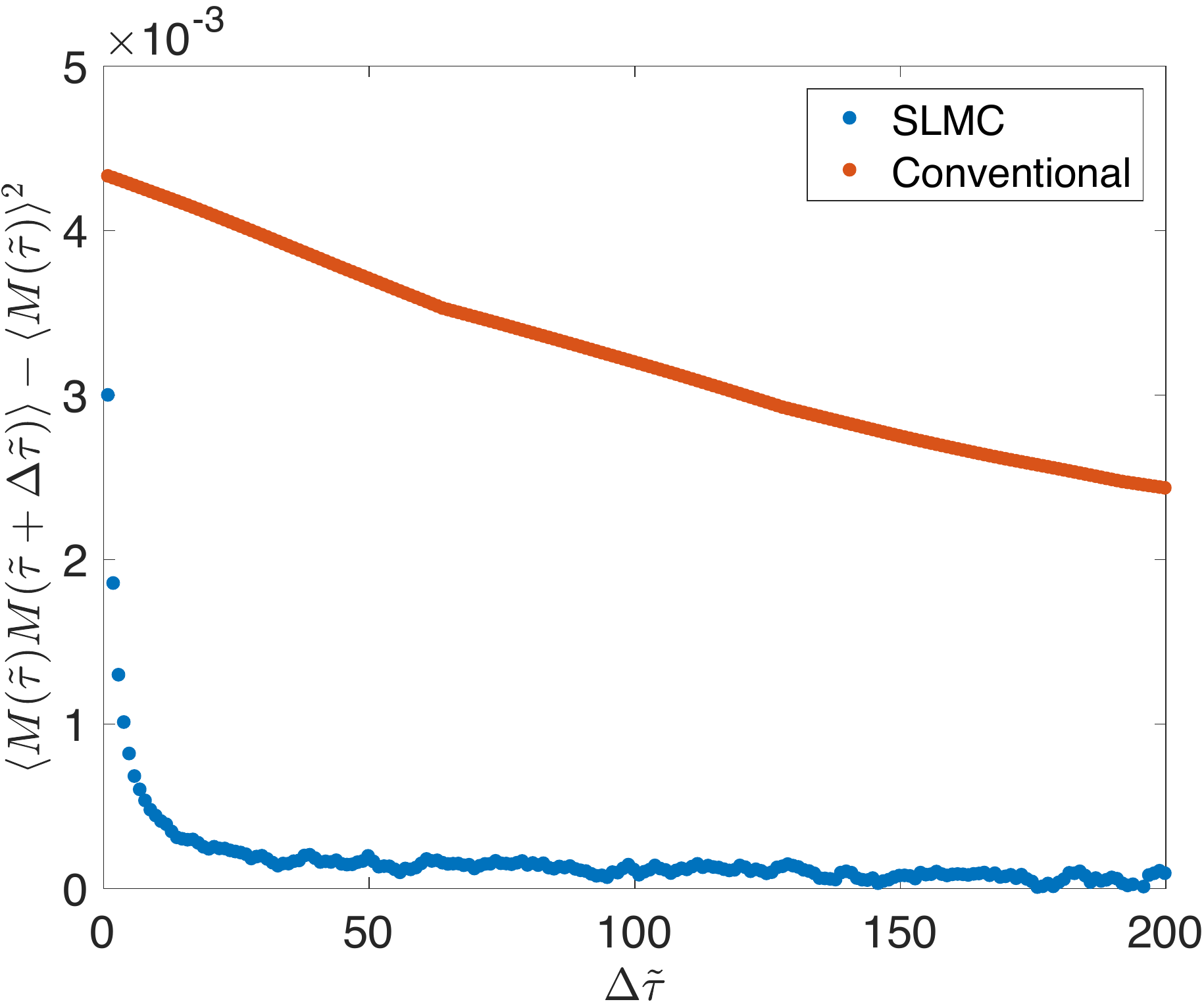}
\caption{(color online) The autocorrelation function of the magnetization with $\tilde{\tau}$ (number of EDs) as the time unit for both the SLMC and the conventional method on the system with $ L=4 $. Here, for SLMC we choose $ l_{\text{c}} =16 \times 4^3$. 
}
\label{fig3}
\end{figure}

Finally we show the actual speedup of our SLMC simulation. Since ED calculations dominate the computational cost in both the conventional and the SLMC method, we use the number of EDs denoted by $\tilde{\tau}$, instead of the number of sweeps $\tau$, to measure the performance. Fig.\ref{fig3} shows the decay of autocorrelation functions with $\tilde{\tau}$ as the time unit for both methods. From these we extract the autocorrelation time $\tilde{\tau}_0$, which precisely characterizes the actual computational cost to obtain two uncorrelated spin configurations. For $L=4$, $\tilde{\tau}_0$ is over 150 times shorter in SLMC than in the conventional method.

\begin{figure}[tbp]
\includegraphics[width=\columnwidth]{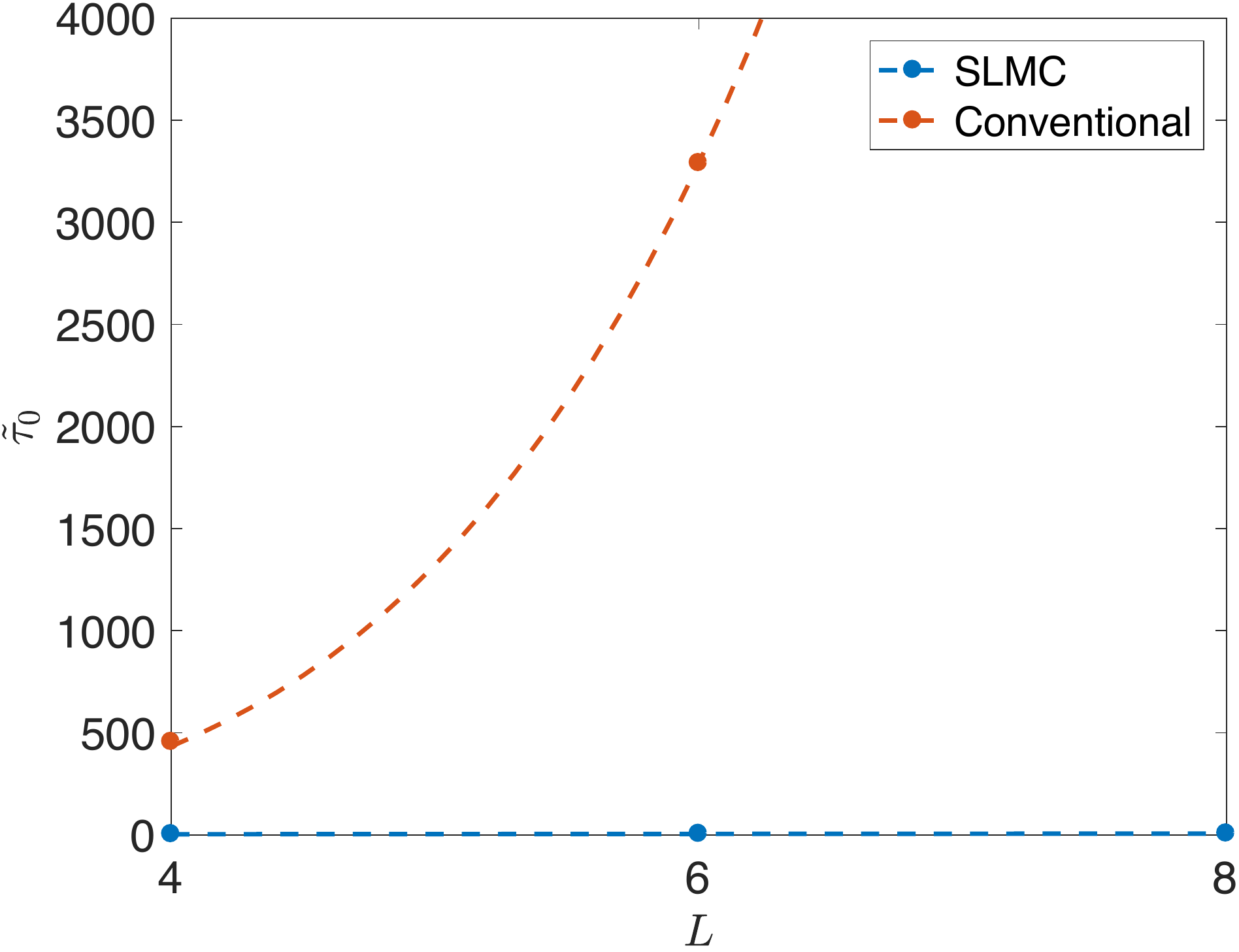}
\caption{(color online) The autocorrelation time $\tilde{\tau}_0$ for both the SLMC and the conventional method on systems with $ L=4,6,8 $. The quadratic red dashed curve for the conventional method is fitted by assuming the dynamical exponent $ z=2 $. 
}
\label{fig4}
\end{figure}

In Fig.\ref{fig4}, we demonstrate the remarkable efficiency of the SLMC method as the system size increases. As discussed previously, in the conventional MC method $\tilde{\tau}_0 \sim O( L^{d+z})$ scales extremely fast with system size $L$ due to the cost of EDs, as confirmed by our expensive numerics for $L=4,6$. In contrast, in SLMC method with an accurate effective model, we expect $\tilde{\tau}_0 \sim O(1) $. Indeed our SLMC simulation shows $\tilde{\tau}_0$ hardly scales with the system size. For $ L=8 $, $\tilde{\tau}_0$ in the conventional method is estimated to be $\sim 10^4 $ from extrapolation based on the scaling, whereas in SLMC it is numerically found to be $\sim 6 $, leading to an enormous speedup by $\sim 10^3 $ times. 

To conclude, we developed a general-purpose, statistically-exact SLMC method with a cumulative update algorithm for simulating fermion systems, whose computational complexity is significantly reduced compared to the conventional MC method. We believe our method holds great promise for solving a wide class of challenging many-body problems of fundamental interest.


{\it Acknowledgement:}
We thank Xiao Yan Xu for helpful discussions. This work is supported by the DOE Office of Basic Energy Sciences, Division of Materials Sciences and Engineering under Award DE-SC0010526. LF is supported partly by the David and Lucile Packard Foundation. JL is supported partly by S3TEC Solid State Solar Thermal Energy Conversion Center, an Energy Frontier Research Center funded by the U.S. Department of Energy (DOE), Office of Science, Basic Energy Sciences (BES), under Award No. DE-SC0001299/DE-FG02-09ER46577. HTS is supported by MIT Alumni Fellowship Fund For Physics. ZYM is supported by the Ministry of Science and Technology (MOST) of China under Grant No. 2016YFA0300502, the National Natural Science Foundation of China (NSFC Grants No. 11421092 and No. 11574359), and the National Thousand-Young-Talents Program of China.

\bibliography{MLMC_Ref}

\begin{thebibliography}{37}%
\makeatletter
\providecommand \@ifxundefined [1]{%
 \@ifx{#1\undefined}
}%
\providecommand \@ifnum [1]{%
 \ifnum #1\expandafter \@firstoftwo
 \else \expandafter \@secondoftwo
 \fi
}%
\providecommand \@ifx [1]{%
 \ifx #1\expandafter \@firstoftwo
 \else \expandafter \@secondoftwo
 \fi
}%
\providecommand \natexlab [1]{#1}%
\providecommand \enquote  [1]{``#1''}%
\providecommand \bibnamefont  [1]{#1}%
\providecommand \bibfnamefont [1]{#1}%
\providecommand \citenamefont [1]{#1}%
\providecommand \href@noop [0]{\@secondoftwo}%
\providecommand \href [0]{\begingroup \@sanitize@url \@href}%
\providecommand \@href[1]{\@@startlink{#1}\@@href}%
\providecommand \@@href[1]{\endgroup#1\@@endlink}%
\providecommand \@sanitize@url [0]{\catcode `\\12\catcode `\$12\catcode
  `\&12\catcode `\#12\catcode `\^12\catcode `\_12\catcode `\%12\relax}%
\providecommand \@@startlink[1]{}%
\providecommand \@@endlink[0]{}%
\providecommand \url  [0]{\begingroup\@sanitize@url \@url }%
\providecommand \@url [1]{\endgroup\@href {#1}{\urlprefix }}%
\providecommand \urlprefix  [0]{URL }%
\providecommand \Eprint [0]{\href }%
\providecommand \doibase [0]{http://dx.doi.org/}%
\providecommand \selectlanguage [0]{\@gobble}%
\providecommand \bibinfo  [0]{\@secondoftwo}%
\providecommand \bibfield  [0]{\@secondoftwo}%
\providecommand \translation [1]{[#1]}%
\providecommand \BibitemOpen [0]{}%
\providecommand \bibitemStop [0]{}%
\providecommand \bibitemNoStop [0]{.\EOS\space}%
\providecommand \EOS [0]{\spacefactor3000\relax}%
\providecommand \BibitemShut  [1]{\csname bibitem#1\endcsname}%
\let\auto@bib@innerbib\@empty
\bibitem [{\citenamefont {Metropolis}\ \emph {et~al.}(1953)\citenamefont
  {Metropolis}, \citenamefont {Rosenbluth}, \citenamefont {Rosenbluth},
  \citenamefont {Teller},\ and\ \citenamefont {Teller}}]{Metropolis1953}%
  \BibitemOpen
  \bibfield  {author} {\bibinfo {author} {\bibfnamefont {N.}~\bibnamefont
  {Metropolis}}, \bibinfo {author} {\bibfnamefont {A.~W.}\ \bibnamefont
  {Rosenbluth}}, \bibinfo {author} {\bibfnamefont {M.~N.}\ \bibnamefont
  {Rosenbluth}}, \bibinfo {author} {\bibfnamefont {A.~H.}\ \bibnamefont
  {Teller}}, \ and\ \bibinfo {author} {\bibfnamefont {E.}~\bibnamefont
  {Teller}},\ }\href {\doibase http://dx.doi.org/10.1063/1.1699114} {\bibfield
  {journal} {\bibinfo  {journal} {J. Chem. Phys.}\ }\textbf {\bibinfo {volume}
  {21}},\ \bibinfo {pages} {1087} (\bibinfo {year} {1953})}\BibitemShut
  {NoStop}%
\bibitem [{\citenamefont {Hastings}(1970)}]{HASTINGS1970}%
  \BibitemOpen
  \bibfield  {author} {\bibinfo {author} {\bibfnamefont {W.~K.}\ \bibnamefont
  {Hastings}},\ }\href {\doibase 10.1093/biomet/57.1.97} {\bibfield  {journal}
  {\bibinfo  {journal} {Biometrika}\ }\textbf {\bibinfo {volume} {57}},\
  \bibinfo {pages} {97} (\bibinfo {year} {1970})}\BibitemShut {NoStop}%
\bibitem [{\citenamefont {Binder}(1995)}]{Binder1995}%
  \BibitemOpen
  \bibinfo {editor} {\bibfnamefont {K.}~\bibnamefont {Binder}},\ ed.,\ \href
  {\doibase 10.1007/3-540-60174-0} {\emph {\bibinfo {title} {The Monte Carlo
  Method in Condensed Matter Physics}}}\ (\bibinfo  {publisher}
  {Springer-Verlag Berlin Heidelberg},\ \bibinfo {year} {1995})\BibitemShut
  {NoStop}%
\bibitem [{\citenamefont {Newman}\ and\ \citenamefont
  {Barkema}(1999)}]{NewmanBarkema1999}%
  \BibitemOpen
  \bibfield  {author} {\bibinfo {author} {\bibfnamefont {M.~E.~J.}\
  \bibnamefont {Newman}}\ and\ \bibinfo {author} {\bibfnamefont {G.~T.}\
  \bibnamefont {Barkema}},\ }\href
  {https://global.oup.com/academic/product/monte-carlo-methods-in-statistical-physics-9780198517979?cc=cn&lang=en&#}
  {\emph {\bibinfo {title} {Monte Carlo Methods in Statistical Physics}}}\
  (\bibinfo  {publisher} {Oxford University Press},\ \bibinfo {year}
  {1999})\BibitemShut {NoStop}%
\bibitem [{\citenamefont {Gubernatis}(2003)}]{Gubernatis2003}%
  \BibitemOpen
  \bibinfo {editor} {\bibfnamefont {J.~E.}\ \bibnamefont {Gubernatis}},\ ed.,\
  \href {http://scitation.aip.org/content/aip/proceeding/aipcp/690} {\emph
  {\bibinfo {title} {THE MONTE CARLO METHOD IN THE PHYSICAL SCIENCES:
  Celebrating the 50th Anniversary of the Metropolis Algorithm}}},\ Vol.\
  \bibinfo {volume} {690},\ \bibinfo {organization} {AIP Conference
  Proceedings}\ (\bibinfo  {publisher} {AIP Publishing},\ \bibinfo {address}
  {Los Alamos, New Mexico (USA)},\ \bibinfo {year} {2003})\BibitemShut
  {NoStop}%
\bibitem [{\citenamefont {Landau}\ and\ \citenamefont
  {Binder}(2005)}]{LandauBinder2005}%
  \BibitemOpen
  \bibfield  {author} {\bibinfo {author} {\bibfnamefont {D.}~\bibnamefont
  {Landau}}\ and\ \bibinfo {author} {\bibfnamefont {K.}~\bibnamefont
  {Binder}},\ }\href {https://books.google.ca/books?id=HbxQxS7tHiYC} {\emph
  {\bibinfo {title} {A Guide to Monte Carlo Simulations in Statistical
  Physics}}}\ (\bibinfo  {publisher} {Cambridge University Press},\ \bibinfo
  {year} {2005})\BibitemShut {NoStop}%
\bibitem [{\citenamefont {Sandvik}(2010)}]{Sandvik2010}%
  \BibitemOpen
  \bibfield  {author} {\bibinfo {author} {\bibfnamefont {A.~W.}\ \bibnamefont
  {Sandvik}},\ }\href {\doibase http://dx.doi.org/10.1063/1.3518900} {\bibfield
   {journal} {\bibinfo  {journal} {AIP Conference Proceedings}\ }\textbf
  {\bibinfo {volume} {1297}},\ \bibinfo {pages} {135} (\bibinfo {year}
  {2010})}\BibitemShut {NoStop}%
\bibitem [{\citenamefont {Liu}\ \emph {et~al.}(2016)\citenamefont {Liu},
  \citenamefont {Qi}, \citenamefont {Meng},\ and\ \citenamefont
  {Fu}}]{liu2016self}%
  \BibitemOpen
  \bibfield  {author} {\bibinfo {author} {\bibfnamefont {J.}~\bibnamefont
  {Liu}}, \bibinfo {author} {\bibfnamefont {Y.}~\bibnamefont {Qi}}, \bibinfo
  {author} {\bibfnamefont {Z.~Y.}\ \bibnamefont {Meng}}, \ and\ \bibinfo
  {author} {\bibfnamefont {L.}~\bibnamefont {Fu}},\ }\href
  {https://arxiv.org/abs/1610.03137} {\bibfield  {journal} {\bibinfo  {journal}
  {arXiv:1610.03137}\ } (\bibinfo {year} {2016})}\BibitemShut {NoStop}%
\bibitem [{\citenamefont {Swendsen}\ and\ \citenamefont
  {Wang}(1987)}]{SwendsenWang1987}%
  \BibitemOpen
  \bibfield  {author} {\bibinfo {author} {\bibfnamefont {R.~H.}\ \bibnamefont
  {Swendsen}}\ and\ \bibinfo {author} {\bibfnamefont {J.-S.}\ \bibnamefont
  {Wang}},\ }\href {\doibase 10.1103/PhysRevLett.58.86} {\bibfield  {journal}
  {\bibinfo  {journal} {Phys. Rev. Lett.}\ }\textbf {\bibinfo {volume} {58}},\
  \bibinfo {pages} {86} (\bibinfo {year} {1987})}\BibitemShut {NoStop}%
\bibitem [{\citenamefont {Wolff}(1989)}]{Wolff1989}%
  \BibitemOpen
  \bibfield  {author} {\bibinfo {author} {\bibfnamefont {U.}~\bibnamefont
  {Wolff}},\ }\href {\doibase 10.1103/PhysRevLett.62.361} {\bibfield  {journal}
  {\bibinfo  {journal} {Phys. Rev. Lett.}\ }\textbf {\bibinfo {volume} {62}},\
  \bibinfo {pages} {361} (\bibinfo {year} {1989})}\BibitemShut {NoStop}%
\bibitem [{\citenamefont {Stratonovich}(1957)}]{stratonovich1957}%
  \BibitemOpen
  \bibfield  {author} {\bibinfo {author} {\bibfnamefont {R.}~\bibnamefont
  {Stratonovich}},\ }in\ \href@noop {} {\emph {\bibinfo {booktitle} {Soviet
  Physics Doklady}}},\ Vol.~\bibinfo {volume} {2}\ (\bibinfo {year} {1957})\
  p.\ \bibinfo {pages} {416}\BibitemShut {NoStop}%
\bibitem [{\citenamefont {Hubbard}(1959)}]{hubbard1959}%
  \BibitemOpen
  \bibfield  {author} {\bibinfo {author} {\bibfnamefont {J.}~\bibnamefont
  {Hubbard}},\ }\href {\doibase 10.1103/PhysRevLett.3.77} {\bibfield  {journal}
  {\bibinfo  {journal} {Phys. Rev. Lett.}\ }\textbf {\bibinfo {volume} {3}},\
  \bibinfo {pages} {77} (\bibinfo {year} {1959})}\BibitemShut {NoStop}%
\bibitem [{\citenamefont {Blankenbecler}\ \emph {et~al.}(1981)\citenamefont
  {Blankenbecler}, \citenamefont {Scalapino},\ and\ \citenamefont
  {Sugar}}]{blankenbecler1981}%
  \BibitemOpen
  \bibfield  {author} {\bibinfo {author} {\bibfnamefont {R.}~\bibnamefont
  {Blankenbecler}}, \bibinfo {author} {\bibfnamefont {D.~J.}\ \bibnamefont
  {Scalapino}}, \ and\ \bibinfo {author} {\bibfnamefont {R.~L.}\ \bibnamefont
  {Sugar}},\ }\href {\doibase 10.1103/PhysRevD.24.2278} {\bibfield  {journal}
  {\bibinfo  {journal} {Phys. Rev. D}\ }\textbf {\bibinfo {volume} {24}},\
  \bibinfo {pages} {2278} (\bibinfo {year} {1981})}\BibitemShut {NoStop}%
\bibitem [{\citenamefont {Hirsch}(1985)}]{hirsch1985}%
  \BibitemOpen
  \bibfield  {author} {\bibinfo {author} {\bibfnamefont {J.~E.}\ \bibnamefont
  {Hirsch}},\ }\href {\doibase 10.1103/PhysRevB.31.4403} {\bibfield  {journal}
  {\bibinfo  {journal} {Phys. Rev. B}\ }\textbf {\bibinfo {volume} {31}},\
  \bibinfo {pages} {4403} (\bibinfo {year} {1985})}\BibitemShut {NoStop}%
\bibitem [{\citenamefont {Hirsch}\ and\ \citenamefont
  {Fye}(1986)}]{hirsch1986}%
  \BibitemOpen
  \bibfield  {author} {\bibinfo {author} {\bibfnamefont {J.~E.}\ \bibnamefont
  {Hirsch}}\ and\ \bibinfo {author} {\bibfnamefont {R.~M.}\ \bibnamefont
  {Fye}},\ }\href {\doibase 10.1103/PhysRevLett.56.2521} {\bibfield  {journal}
  {\bibinfo  {journal} {Physical review letters}\ }\textbf {\bibinfo {volume}
  {56}},\ \bibinfo {pages} {2521} (\bibinfo {year} {1986})}\BibitemShut
  {NoStop}%
\bibitem [{\citenamefont {White}\ \emph {et~al.}(1989)\citenamefont {White},
  \citenamefont {Scalapino}, \citenamefont {Sugar}, \citenamefont {Loh},
  \citenamefont {Gubernatis},\ and\ \citenamefont {Scalettar}}]{white1989}%
  \BibitemOpen
  \bibfield  {author} {\bibinfo {author} {\bibfnamefont {S.~R.}\ \bibnamefont
  {White}}, \bibinfo {author} {\bibfnamefont {D.~J.}\ \bibnamefont
  {Scalapino}}, \bibinfo {author} {\bibfnamefont {R.~L.}\ \bibnamefont
  {Sugar}}, \bibinfo {author} {\bibfnamefont {E.~Y.}\ \bibnamefont {Loh}},
  \bibinfo {author} {\bibfnamefont {J.~E.}\ \bibnamefont {Gubernatis}}, \ and\
  \bibinfo {author} {\bibfnamefont {R.~T.}\ \bibnamefont {Scalettar}},\ }\href
  {\doibase 10.1103/PhysRevB.40.506} {\bibfield  {journal} {\bibinfo  {journal}
  {Physical Review B}\ }\textbf {\bibinfo {volume} {40}},\ \bibinfo {pages}
  {506} (\bibinfo {year} {1989})}\BibitemShut {NoStop}%
\bibitem [{\citenamefont {Trotter}(1959)}]{trotter1959}%
  \BibitemOpen
  \bibfield  {author} {\bibinfo {author} {\bibfnamefont {H.~F.}\ \bibnamefont
  {Trotter}},\ }\href@noop {} {\bibfield  {journal} {\bibinfo  {journal}
  {Proceedings of the American Mathematical Society}\ }\textbf {\bibinfo
  {volume} {10}},\ \bibinfo {pages} {545} (\bibinfo {year} {1959})}\BibitemShut
  {NoStop}%
\bibitem [{\citenamefont {Suzuki}(1976)}]{suzuki1976}%
  \BibitemOpen
  \bibfield  {author} {\bibinfo {author} {\bibfnamefont {M.}~\bibnamefont
  {Suzuki}},\ }\href@noop {} {\bibfield  {journal} {\bibinfo  {journal}
  {Communications in Mathematical Physics}\ }\textbf {\bibinfo {volume} {51}},\
  \bibinfo {pages} {183} (\bibinfo {year} {1976})}\BibitemShut {NoStop}%
\bibitem [{\citenamefont {Berg}\ \emph {et~al.}(2012)\citenamefont {Berg},
  \citenamefont {Metlitski},\ and\ \citenamefont {Sachdev}}]{Erez2012}%
  \BibitemOpen
  \bibfield  {author} {\bibinfo {author} {\bibfnamefont {E.}~\bibnamefont
  {Berg}}, \bibinfo {author} {\bibfnamefont {M.~A.}\ \bibnamefont {Metlitski}},
  \ and\ \bibinfo {author} {\bibfnamefont {S.}~\bibnamefont {Sachdev}},\ }\href
  {\doibase 10.1126/science.1227769} {\bibfield  {journal} {\bibinfo  {journal}
  {Science}\ }\textbf {\bibinfo {volume} {338}},\ \bibinfo {pages} {1606}
  (\bibinfo {year} {2012})}\BibitemShut {NoStop}%
\bibitem [{\citenamefont {Lederer}\ \emph {et~al.}(2015)\citenamefont
  {Lederer}, \citenamefont {Schattner}, \citenamefont {Berg},\ and\
  \citenamefont {Kivelson}}]{Lederer2015}%
  \BibitemOpen
  \bibfield  {author} {\bibinfo {author} {\bibfnamefont {S.}~\bibnamefont
  {Lederer}}, \bibinfo {author} {\bibfnamefont {Y.}~\bibnamefont {Schattner}},
  \bibinfo {author} {\bibfnamefont {E.}~\bibnamefont {Berg}}, \ and\ \bibinfo
  {author} {\bibfnamefont {S.~A.}\ \bibnamefont {Kivelson}},\ }\href {\doibase
  10.1103/PhysRevLett.114.097001} {\bibfield  {journal} {\bibinfo  {journal}
  {Phys. Rev. Lett.}\ }\textbf {\bibinfo {volume} {114}},\ \bibinfo {pages}
  {097001} (\bibinfo {year} {2015})}\BibitemShut {NoStop}%
\bibitem [{\citenamefont {Li}\ \emph {et~al.}(2015)\citenamefont {Li},
  \citenamefont {Wang}, \citenamefont {Yao},\ and\ \citenamefont
  {Lee}}]{li2015a}%
  \BibitemOpen
  \bibfield  {author} {\bibinfo {author} {\bibfnamefont {Z.-X.}\ \bibnamefont
  {Li}}, \bibinfo {author} {\bibfnamefont {F.}~\bibnamefont {Wang}}, \bibinfo
  {author} {\bibfnamefont {H.}~\bibnamefont {Yao}}, \ and\ \bibinfo {author}
  {\bibfnamefont {D.-H.}\ \bibnamefont {Lee}},\ }\href@noop {} {\bibfield
  {journal} {\bibinfo  {journal} {arXiv:1512.04541}\ } (\bibinfo {year}
  {2015})}\BibitemShut {NoStop}%
\bibitem [{\citenamefont {Schattner}\ \emph
  {et~al.}(2016{\natexlab{a}})\citenamefont {Schattner}, \citenamefont
  {Gerlach}, \citenamefont {Trebst},\ and\ \citenamefont
  {Berg}}]{Schattner2016prl}%
  \BibitemOpen
  \bibfield  {author} {\bibinfo {author} {\bibfnamefont {Y.}~\bibnamefont
  {Schattner}}, \bibinfo {author} {\bibfnamefont {M.~H.}\ \bibnamefont
  {Gerlach}}, \bibinfo {author} {\bibfnamefont {S.}~\bibnamefont {Trebst}}, \
  and\ \bibinfo {author} {\bibfnamefont {E.}~\bibnamefont {Berg}},\ }\href
  {\doibase 10.1103/PhysRevLett.117.097002} {\bibfield  {journal} {\bibinfo
  {journal} {Phys. Rev. Lett.}\ }\textbf {\bibinfo {volume} {117}},\ \bibinfo
  {pages} {097002} (\bibinfo {year} {2016}{\natexlab{a}})}\BibitemShut
  {NoStop}%
\bibitem [{\citenamefont {Schattner}\ \emph
  {et~al.}(2016{\natexlab{b}})\citenamefont {Schattner}, \citenamefont
  {Lederer}, \citenamefont {Kivelson},\ and\ \citenamefont
  {Berg}}]{Schattner2016prx}%
  \BibitemOpen
  \bibfield  {author} {\bibinfo {author} {\bibfnamefont {Y.}~\bibnamefont
  {Schattner}}, \bibinfo {author} {\bibfnamefont {S.}~\bibnamefont {Lederer}},
  \bibinfo {author} {\bibfnamefont {S.~A.}\ \bibnamefont {Kivelson}}, \ and\
  \bibinfo {author} {\bibfnamefont {E.}~\bibnamefont {Berg}},\ }\href {\doibase
  10.1103/PhysRevX.6.031028} {\bibfield  {journal} {\bibinfo  {journal} {Phys.
  Rev. X}\ }\textbf {\bibinfo {volume} {6}},\ \bibinfo {pages} {031028}
  (\bibinfo {year} {2016}{\natexlab{b}})}\BibitemShut {NoStop}%
\bibitem [{\citenamefont {Xu}\ \emph {et~al.}(2016)\citenamefont {Xu},
  \citenamefont {Beach}, \citenamefont {Sun}, \citenamefont {Assaad},\ and\
  \citenamefont {Meng}}]{xu2016arXiv}%
  \BibitemOpen
  \bibfield  {author} {\bibinfo {author} {\bibfnamefont {X.~Y.}\ \bibnamefont
  {Xu}}, \bibinfo {author} {\bibfnamefont {K.}~\bibnamefont {Beach}}, \bibinfo
  {author} {\bibfnamefont {K.}~\bibnamefont {Sun}}, \bibinfo {author}
  {\bibfnamefont {F.}~\bibnamefont {Assaad}}, \ and\ \bibinfo {author}
  {\bibfnamefont {Z.~Y.}\ \bibnamefont {Meng}},\ }\href@noop {} {\bibfield
  {journal} {\bibinfo  {journal} {arXiv:1602.07150}\ } (\bibinfo {year}
  {2016})}\BibitemShut {NoStop}%
\bibitem [{\citenamefont {Zener}(1951)}]{zener1951}%
  \BibitemOpen
  \bibfield  {author} {\bibinfo {author} {\bibfnamefont {C.}~\bibnamefont
  {Zener}},\ }\href {\doibase 10.1103/PhysRev.82.403} {\bibfield  {journal}
  {\bibinfo  {journal} {Physical Review}\ }\textbf {\bibinfo {volume} {82}},\
  \bibinfo {pages} {403} (\bibinfo {year} {1951})}\BibitemShut {NoStop}%
\bibitem [{\citenamefont {Anderson}\ and\ \citenamefont
  {Hasegawa}(1955)}]{anderson1955}%
  \BibitemOpen
  \bibfield  {author} {\bibinfo {author} {\bibfnamefont {P.~W.}\ \bibnamefont
  {Anderson}}\ and\ \bibinfo {author} {\bibfnamefont {H.}~\bibnamefont
  {Hasegawa}},\ }\href {\doibase 10.1103/PhysRev.100.675} {\bibfield  {journal}
  {\bibinfo  {journal} {Phys. Rev.}\ }\textbf {\bibinfo {volume} {100}},\
  \bibinfo {pages} {675} (\bibinfo {year} {1955})}\BibitemShut {NoStop}%
\bibitem [{\citenamefont {de~Gennes}(1960)}]{de1960}%
  \BibitemOpen
  \bibfield  {author} {\bibinfo {author} {\bibfnamefont {P.~G.}\ \bibnamefont
  {de~Gennes}},\ }\href {\doibase 10.1103/PhysRev.118.141} {\bibfield
  {journal} {\bibinfo  {journal} {Physical Review}\ }\textbf {\bibinfo {volume}
  {118}},\ \bibinfo {pages} {141} (\bibinfo {year} {1960})}\BibitemShut
  {NoStop}%
\bibitem [{\citenamefont {Xu}\ \emph {et~al.}()\citenamefont {Xu},
  \citenamefont {Qi}, \citenamefont {Liu}, \citenamefont {Fu},\ and\
  \citenamefont {Meng}}]{xu2016}%
  \BibitemOpen
  \bibfield  {author} {\bibinfo {author} {\bibfnamefont {X.-Y.}\ \bibnamefont
  {Xu}}, \bibinfo {author} {\bibfnamefont {Y.}~\bibnamefont {Qi}}, \bibinfo
  {author} {\bibfnamefont {J.}~\bibnamefont {Liu}}, \bibinfo {author}
  {\bibfnamefont {L.}~\bibnamefont {Fu}}, \ and\ \bibinfo {author}
  {\bibfnamefont {Z.-Y.}\ \bibnamefont {Meng}},\ }\href@noop {} {}\bibinfo
  {note} {In preparation}\BibitemShut {NoStop}%
\bibitem [{\citenamefont {Motome}\ and\ \citenamefont
  {Furukawa}(1999)}]{motome1999}%
  \BibitemOpen
  \bibfield  {author} {\bibinfo {author} {\bibfnamefont {Y.}~\bibnamefont
  {Motome}}\ and\ \bibinfo {author} {\bibfnamefont {N.}~\bibnamefont
  {Furukawa}},\ }\href@noop {} {\bibfield  {journal} {\bibinfo  {journal}
  {Journal of the Physical Society of Japan}\ }\textbf {\bibinfo {volume}
  {68}},\ \bibinfo {pages} {3853} (\bibinfo {year} {1999})}\BibitemShut
  {NoStop}%
\bibitem [{\citenamefont {Alonso}\ \emph {et~al.}(2001)\citenamefont {Alonso},
  \citenamefont {Fern{\'a}ndez}, \citenamefont {Guinea}, \citenamefont
  {Laliena},\ and\ \citenamefont {Mart{\'\i}n-Mayor}}]{alonso2001}%
  \BibitemOpen
  \bibfield  {author} {\bibinfo {author} {\bibfnamefont {J.}~\bibnamefont
  {Alonso}}, \bibinfo {author} {\bibfnamefont {L.}~\bibnamefont
  {Fern{\'a}ndez}}, \bibinfo {author} {\bibfnamefont {F.}~\bibnamefont
  {Guinea}}, \bibinfo {author} {\bibfnamefont {V.}~\bibnamefont {Laliena}}, \
  and\ \bibinfo {author} {\bibfnamefont {V.}~\bibnamefont
  {Mart{\'\i}n-Mayor}},\ }\href {\doibase
  http://dx.doi.org/10.1016/S0550-3213(00)00681-7} {\bibfield  {journal}
  {\bibinfo  {journal} {Nuclear Physics B}\ }\textbf {\bibinfo {volume}
  {596}},\ \bibinfo {pages} {587} (\bibinfo {year} {2001})}\BibitemShut
  {NoStop}%
\bibitem [{\citenamefont {Alvarez}\ \emph {et~al.}(2005)\citenamefont
  {Alvarez}, \citenamefont {{\c{S}}en}, \citenamefont {Furukawa}, \citenamefont
  {Motome},\ and\ \citenamefont {Dagotto}}]{alvarez2005}%
  \BibitemOpen
  \bibfield  {author} {\bibinfo {author} {\bibfnamefont {G.}~\bibnamefont
  {Alvarez}}, \bibinfo {author} {\bibfnamefont {C.}~\bibnamefont {{\c{S}}en}},
  \bibinfo {author} {\bibfnamefont {N.}~\bibnamefont {Furukawa}}, \bibinfo
  {author} {\bibfnamefont {Y.}~\bibnamefont {Motome}}, \ and\ \bibinfo {author}
  {\bibfnamefont {E.}~\bibnamefont {Dagotto}},\ }\href {\doibase
  http://dx.doi.org/10.1016/j.cpc.2005.02.001} {\bibfield  {journal} {\bibinfo
  {journal} {Computer Physics Communications}\ }\textbf {\bibinfo {volume}
  {168}},\ \bibinfo {pages} {32} (\bibinfo {year} {2005})}\BibitemShut
  {NoStop}%
\bibitem [{\citenamefont {Kumar}\ and\ \citenamefont
  {Majumdar}(2006)}]{Kumar2006}%
  \BibitemOpen
  \bibfield  {author} {\bibinfo {author} {\bibfnamefont {S.}~\bibnamefont
  {Kumar}}\ and\ \bibinfo {author} {\bibfnamefont {P.}~\bibnamefont
  {Majumdar}},\ }\href {\doibase 10.1140/epjb/e2006-00173-2} {\bibfield
  {journal} {\bibinfo  {journal} {The European Physical Journal B}\ }\textbf
  {\bibinfo {volume} {50}},\ \bibinfo {pages} {571} (\bibinfo {year}
  {2006})}\BibitemShut {NoStop}%
\bibitem [{\citenamefont {Kumar}\ and\ \citenamefont {van~den
  Brink}(2010)}]{Jeroen2010}%
  \BibitemOpen
  \bibfield  {author} {\bibinfo {author} {\bibfnamefont {S.}~\bibnamefont
  {Kumar}}\ and\ \bibinfo {author} {\bibfnamefont {J.}~\bibnamefont {van~den
  Brink}},\ }\href {\doibase 10.1103/PhysRevLett.105.216405} {\bibfield
  {journal} {\bibinfo  {journal} {Phys. Rev. Lett.}\ }\textbf {\bibinfo
  {volume} {105}},\ \bibinfo {pages} {216405} (\bibinfo {year}
  {2010})}\BibitemShut {NoStop}%
\bibitem [{\citenamefont {Mukherjee}\ \emph {et~al.}(2015)\citenamefont
  {Mukherjee}, \citenamefont {Patel}, \citenamefont {Bishop},\ and\
  \citenamefont {Dagotto}}]{mukherjee2015}%
  \BibitemOpen
  \bibfield  {author} {\bibinfo {author} {\bibfnamefont {A.}~\bibnamefont
  {Mukherjee}}, \bibinfo {author} {\bibfnamefont {N.~D.}\ \bibnamefont
  {Patel}}, \bibinfo {author} {\bibfnamefont {C.}~\bibnamefont {Bishop}}, \
  and\ \bibinfo {author} {\bibfnamefont {E.}~\bibnamefont {Dagotto}},\ }\href
  {\doibase 10.1103/PhysRevE.91.063303} {\bibfield  {journal} {\bibinfo
  {journal} {Phys. Rev. E}\ }\textbf {\bibinfo {volume} {91}},\ \bibinfo
  {pages} {063303} (\bibinfo {year} {2015})}\BibitemShut {NoStop}%
\bibitem [{\citenamefont {Ruderman}\ and\ \citenamefont
  {Kittel}(1954)}]{ruderman1954}%
  \BibitemOpen
  \bibfield  {author} {\bibinfo {author} {\bibfnamefont {M.~A.}\ \bibnamefont
  {Ruderman}}\ and\ \bibinfo {author} {\bibfnamefont {C.}~\bibnamefont
  {Kittel}},\ }\href {\doibase 10.1103/PhysRev.96.99} {\bibfield  {journal}
  {\bibinfo  {journal} {Phys. Rev.}\ }\textbf {\bibinfo {volume} {96}},\
  \bibinfo {pages} {99} (\bibinfo {year} {1954})}\BibitemShut {NoStop}%
\bibitem [{\citenamefont {Kasuya}(1956)}]{kasuya1956}%
  \BibitemOpen
  \bibfield  {author} {\bibinfo {author} {\bibfnamefont {T.}~\bibnamefont
  {Kasuya}},\ }\href {\doibase 10.1143/PTP.16.45} {\bibfield  {journal}
  {\bibinfo  {journal} {Progress of Theoretical Physics}\ }\textbf {\bibinfo
  {volume} {16}},\ \bibinfo {pages} {45} (\bibinfo {year} {1956})}\BibitemShut
  {NoStop}%
\bibitem [{\citenamefont {Yosida}(1957)}]{yosida1957}%
  \BibitemOpen
  \bibfield  {author} {\bibinfo {author} {\bibfnamefont {K.}~\bibnamefont
  {Yosida}},\ }\href {\doibase 10.1103/PhysRev.106.893} {\bibfield  {journal}
  {\bibinfo  {journal} {Phys. Rev.}\ }\textbf {\bibinfo {volume} {106}},\
  \bibinfo {pages} {893} (\bibinfo {year} {1957})}\BibitemShut {NoStop}%
\end{thebibliography}%

\end{document}